\documentclass[twocolumn,showpacs,preprintnumbers,amsmath,amssymb, prb]{revtex4}

\usepackage{graphicx}
\usepackage{dcolumn}
\usepackage{bm}

\begin{document}


\title{Fano effect in a Josephson junction with a quantum dot}

\author{Kentaro Osawa}
\email{osawa@kh.phys.waseda.ac.jp}
\affiliation{Department of Physics, Waseda University, Okubo, Shinjuku, Tokyo 169-8555, Japan.}
\author{Nobuhiko Yokoshi}
\affiliation{Nanotechnology Research Institute, AIST, Tukuba 305-8568, Japan}
\affiliation{CREST(JST), Saitama 332-0012, Japan}
\author{Susumu Kurihara}
\affiliation{Department of Physics, Waseda University, Okubo, Shinjuku, Tokyo 169-8555, Japan.}
\homepage{http://www.kh.phys.waseda.ac.jp/}
\date{\today}

\begin{abstract}
We theoretically investigate the Fano effect in dc Josephson current at the absolute zero of temperature. The system under consideration is a double-path Josephson junction in which one path is through an insulating barrier and the other one is through a quantum dot (QD). Here the Kondo temperature is assumed to be much smaller than the superconducting gap, and the Coulomb interaction inside the QD is treated by the Hartree-Fock approximation. It is shown that the Josephson critical current exhibits an asymmetric resonance against the QD energy level. This behavior is caused by the interference between the two tunneling processes between the superconductors; the direct tunneling across the insulating barrier and the resonant one through the QD. Moreover, we find that the Josephson critical current changes its sign around the resonance when the Coulomb interaction is sufficiently strong. Our results suggest that $0$-$\pi$ transition is induced by the cooperation of the Fano effect and the Coulomb interaction inside the QD.
\end{abstract}

\maketitle

\section{\label{1}Introduction} 
In the last decade, transport properties of quantum dot (QD) systems have been extensively studied. These studies are motivated not only by their potential applications~\cite{QC,QDL,SET} but also by fundamental interests in many body problems and interference phenomena, such as the Kondo effect~\cite{Kondo,Kondo2,Kondo3,Kondo4} and the Fano effect~\cite{Fano,Fano1,Fano2,Fano3,Kobayashi}. 

Kobayashi {\it et al}. observed the Fano effect using a QD embedded in an Aharonov-Bohm interferometer~\cite{Kobayashi}. In this system, there are two transmission processes between the source and drain electrodes. One is through the continuum in the arm, and the other is through the discrete energy level of the QD. The interference between these processes results in an asymmetric resonance in the differential conductance. Thereafter there have been many studies on the Fano effect in QD systems, but most of these studies have focused on normal metal systems~\cite{Fano1,Fano2,Fano3}. Thus, the Fano effect in superconducting systems~\cite{senko, Fano4} is less understood.

Meanwhile, many authors have investigated superconductor/quantum dot/superconductor (S/QD/S) junctions~\cite{SDS1,SDS2,SDS3,SDS4,SDS5, SDS6,SDS7,SDS8,SDS9}. In these systems, the ratio of the Kondo temperature $T_{{\rm K}}$ to the superconducting gap $\Delta$ is a key parameter. In the strong coupling limit $T_{{\rm K}} \gg \Delta$, the Kondo effect survives even in the presence of the superconductivity; a Cooper pair is broken in order to screen the localized spin in the QD. On the other hand, in the weak coupling limit $T_{{\rm K}} \ll \Delta$, the Kondo effect is negligible because a strongly bound Cooper pair cannot be broken. Then, the Cooper pair feels the localized magnetic moment in the QD. Under this situation, when the Coulomb interaction is strong inside the QD, so-called $0$-$\pi$ transition occurs~\cite{pi1,pi2,pi3,pi HF1,pi HF2}. It means that the dependence of the Josephson current on the phase difference $\theta$ changes from $I=|I_{{\rm c}}| \sin \theta$ to $I=|I_{{\rm c}}| \sin (\theta+\pi)=-|I_{{\rm c}}| \sin \theta$, i.e., the critical current becomes negative. 

Recently, Zhang studied the Fano effect in a Josephson junction with a QD coupled in parallel~\cite{senko}. The calculations were done within the finite $U$ slave boson mean field theory for $T_{{\rm K}} > \Delta$. It was concluded that the Fano effect caused $0$-$\pi$ transition in that regime. 

\begin{figure}[t]
\begin{center}
\includegraphics[width=60mm]{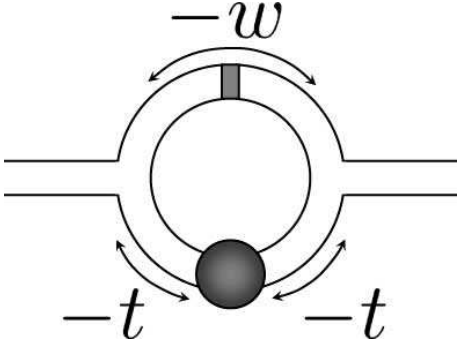}
\caption{\label{fig1} Schematic illustration of the system under consideration. A S/QD/S junction and a conventional Josephson junction are connected in parallel.}
\end{center}
\end{figure}

In this paper, we calculate the Josephson current through the similar system as in Ref.~\onlinecite{senko}, but for $T_{{\rm K}} \ll \Delta$. The purpose of this study is to examine whether the $0$-$\pi$ transition caused by the Fano effect occurs or not in this regime. The system we consider is a double-path Josephson junction; a S/QD/S junction and a conventional Josephson junction are connected in parallel (see Fig.~\ref{fig1}). Here we employ the Hartree-Fock approximation (HFA) in treating the Coulomb interaction inside the QD. 
It is found that the critical current exhibits the characteristic Fano-like dependence on the QD energy level. We also show that the critical current near the resonance can change its sign under the strong Coulomb interaction. This means that the combination of the Fano effect and the Coulomb interaction effect causes $0$-$\pi$ transition.
Finally, by using a simple perturbative approach with respect to the electron tunneling, we discuss the physical origin of the transition and estimate the parameter regime where the system behaves as a $\pi$-junction.

The organization of this paper is as follows. In the following section, we introduce the model Hamiltonian and present the details of our calculation. In Sec.~I\hspace{-.1em}I\hspace{-.1em}I, we show the results for the non-interacting case and the interacting case respectively. Discussions are given in Sec.~I\hspace{-.1em}V with the results of the perturbation expansion. Section~I\hspace{-.1em}V is devoted to a brief summary. We also present the results obtained by the Hubbard-I approximation \cite{Hubbard} in the Appendix.
\section{\label{2}Model and Formalism}
In order to describe the system shown in Fig. \ref{fig1}, we use the following Hamiltonian:
\begin{eqnarray} 
&&{\cal H}={\cal H}_{{\rm L}}+{\cal H}_{{\rm R}}+{\cal H}_{{\rm D}}+{\cal H}_{{\rm T}},
\\ 
&&{\cal H}_{\beta}=\sum_{k,\sigma} \xi_{k} c_{\beta k,\sigma}^\dagger c_{\beta k,\sigma}
\nonumber \\
&& \ \ \ \ \ \ \ \ -\sum_{k}(\Delta {\rm e}^{i\theta_{\beta}} c_{\beta k,\uparrow}^\dagger c_{\beta-k,\downarrow}^\dagger+h.c.),
\\
&&{\cal H}_{{\rm D}}=\sum_{\sigma} \epsilon_{{\rm d}} d_{\sigma}^\dagger d_{\sigma}+U n_{ \uparrow} n_{\downarrow},
\\
&&{\cal H}_{{\rm T}}  =
  -t \sum_{k,\sigma} \Bigl(
c_{{\rm L}k,\sigma}^\dagger d_{\sigma}
+ c_{{\rm R}k,\sigma}^\dagger d_{\sigma} +h.c.
\Bigr)
\nonumber \\
&&~~~~~~~~~~~~~~~~~~
  -w\sum_{k,k',\sigma}
\Bigl(
 c_{{\rm R}k',\sigma}^\dagger c_{{\rm L}k,\sigma}+h.c.
\Bigr),
\label{tunnel}
\end{eqnarray}
where $\beta={\rm L},{\rm R}$. ${\cal H}_{{\rm L}}$ (${\cal H}_{{\rm R}}$) describes the left (right) superconductor with the s-wave order parameter $\Delta {\rm e}^{i \theta_{{\rm L}}}$ ($\Delta {\rm e}^{i \theta_{{\rm R}}}$). $c_{\beta k,\sigma}$ is the annihilation operator for electrons with spin $\sigma$ and wave vector $k$ in the superconductor $\beta$. ${\cal H}_{{\rm D}}$ represents the QD with the discrete energy level $\epsilon_{{\rm d}}$ and the Coulomb interaction $U$, where $d_{\sigma}$ annihilates an electron in the QD and $n_{\sigma}=d_{\sigma}^{\dagger} d_{\sigma}$. The electron tunneling is described by ${\cal H}_{{\rm T}}$, in which $t$ is the tunneling amplitude between the superconductors and the QD, and $w$ is the direct tunneling amplitude across the insulating barrier. 

The Josephson current is simply given by 
\begin{eqnarray}
I=e \langle \dot{N}_{L} \rangle=I_{t}+I_{w},
\end{eqnarray}
where
\begin{eqnarray}
&&I_{t}=-\frac{et}{2 \pi \hbar} \sum_{k} \int {\rm d}\omega~{\rm tr} \left[ \hat{G}_{{\rm d},{\rm L}k}^{<}(\omega)- \hat{G}_{{\rm L}k,{\rm d}}^{<}(\omega) \right],
\\
&&I_{w}=-\frac{ew}{2 \pi \hbar} \sum_{k,k'} \int {\rm d}\omega~{\rm tr} \left[ \hat{G}_{{\rm R}k,{\rm L}k'}^{<}(\omega)- \hat{G}_{{\rm L}k',{\rm R}k}^{<}(\omega)  \right]. \nonumber \\
\end{eqnarray}
Here we introduced the lesser Green functions, which are defined by
\begin{eqnarray}
&&\hat{G}_{\beta k,\beta ' k'}^{<}(t,t')= \\ \nonumber
&&i \left(
\begin{array}{ccc}
\langle c_{\beta ' k',\uparrow}^{\dagger}(t') c_{\beta k,\uparrow}(t)  \rangle &
\langle c_{\beta ' -k',\downarrow}(t') c_{\beta k,\uparrow}(t)  \rangle \\
\langle c_{\beta ' k',\uparrow}^{\dagger}(t') c_{\beta -k,\downarrow}^{\dagger}(t)  \rangle &
\langle c_{\beta ' -k',\downarrow}(t') c_{\beta -k,\downarrow}^{\dagger}(t)  \rangle \\
\end{array}
\right),
\end{eqnarray}
\begin{eqnarray}
&&\hat{G}_{{\rm d},\beta k}^{<}(t,t')= \\ \nonumber
&&i \left(
\begin{array}{ccc}
\langle c_{\beta k,\uparrow}^{\dagger}(t') d_{\uparrow}(t)  \rangle &
\langle c_{\beta -k,\downarrow}(t') d_{\uparrow}(t)  \rangle \\
\langle c_{\beta k,\uparrow}^{\dagger}(t') d_{\downarrow}^{\dagger}(t)  \rangle &
\langle c_{\beta -k,\downarrow}(t') d_{\downarrow}^{\dagger}(t)  \rangle \\
\end{array}
\right).
\end{eqnarray}
In equilibrium state, the lesser Green function is expressed with the advanced and retarded Green functions as
\begin{eqnarray}
\hat{G}_{i,j}^{<}(\omega)=f(\omega) \left( \hat{G}_{i,j}^{a}(\omega) - \hat{G}_{i,j}^{r}(\omega) \right),
\end{eqnarray}
where $f(\omega)$ is the Fermi distribution function. $\hat{G}_{i,j}^{a}$ ($\hat{G}_{i,j}^{r}$) is the advanced (retarded) Green function and $i,j=\{{\rm L}k,{\rm R}k,{\rm d}\}$. Since $( \hat{G}_{i,j}^{a} )^{\dagger}=\hat{G}_{j,i}^{r}$, all we have to calculate is the retarded Green function.
With use of the spinor field operators
\begin{eqnarray}
\psi_{\beta k}=\left( 
\begin{array}{ccc}
 c_{\beta k,\uparrow}  \\
 c_{\beta -k,\downarrow}^{\dagger}  \\
\end{array} 
\right),
\ 
\psi_{{\rm d}}=\left( 
\begin{array}{ccc}
 d_{\uparrow}  \\
 d_{\downarrow}^{\dagger}  \\
\end{array} 
\right),
\end{eqnarray}
we express the retarded Green function as
\begin{eqnarray}
\hat{G}_{i,j}^{r}(\omega)=\langle \langle \psi_{i} ;\psi_{j}^\dagger \rangle \rangle_{\omega}.
\end{eqnarray}

In calculating the Green functions, we employ the equation of motion method. The equation of motion is given by
\begin{eqnarray}
\omega \langle \langle \psi_{i} ;\psi_{j}^\dagger \rangle \rangle_{\omega} = \langle \{\psi_{i},\psi_{j}^\dagger\} \rangle + \langle \langle \psi_{i} ;[{\cal H},\psi_{j}^\dagger] \rangle \rangle_{\omega}.
\end{eqnarray}
It is straightforward to find that all the Green functions are expressed in terms of the dot Green function as
\begin{eqnarray}
&&\hat{G}_{{\rm d} \beta}^{r} = \hat{G}_{{\rm dd}}^{r} \hat{T} \left[ \hat{G}_{\beta \beta}^{(0)r} + \hat{G}_{\bar{\beta} \beta}^{(0)r} \right], \label{Gdb}
\\
&&\hat{G}_{\beta {\rm d}}^{r} = \left[ \hat{G}_{\beta \beta}^{(0)r} + \hat{G}_{\beta \bar{\beta}}^{(0)r} \right] \hat{T} \hat{G}_{{\rm dd}}^{r},
\\
&&\hat{G}_{\beta \bar{\beta}}^{r} = \hat{G}_{\beta \bar{\beta}}^{(0)r} + \left[ \hat{G}_{\beta \beta}^{(0)r} + \hat{G}_{\beta \bar{\beta}}^{(0)r} \right] \hat{T} \hat{G}_{{\rm d} \bar{\beta}}^{r},
\end{eqnarray}
where we have taken the summation over wave vectors in the right hand sides. Here $\bar{{\rm L}}={\rm R}$, $\bar{{\rm R}}={\rm L}$ and $\hat{T}=-t \hat{\sigma}_{{\rm z}}$. $\hat{G}_{\beta \beta}^{(0)r}$ and $\hat{G}_{\beta \bar{\beta}}^{(0)r}$ are the Green functions for $\hat{T}=0$ which are expressed as
\begin{eqnarray}
&&\hat{G}_{\beta \beta}^{(0)r} = \left[(\hat{g}_{\beta}^{r})^{-1} - \hat{W} \hat{g}_{\bar{\beta}}^{r} \hat{W} \right]^{-1},
\\
&&\hat{G}_{\beta \bar{\beta}}^{(0)r} = \hat{g}_{\beta}^{r} \hat{W} \hat{G}_{\bar{\beta} \bar{\beta}}^{(0)r},
\end{eqnarray}
where $\hat{W}=- w\hat{\sigma}_{{\rm z}}$. The unperturbed Green function of the superconductor $\beta$ is given by
\begin{eqnarray}
\hat{g}_{\beta}^{r}
=\frac{\pi \nu}{\sqrt[]{\Delta^{2} - \omega^{2}}} \left( \begin{array}{ccc}
 -\omega  &
 \Delta {\rm e}^{i \theta_{\beta}} \\
 \Delta {\rm e}^{-i \theta_{\beta}}  &
 -\omega \\
\end{array} 
\right),
\end{eqnarray}
where $\nu$ is the normal density of states at the Fermi level. Then, what we have to know in calculating the Josephson current is only the dot Green function $\hat{G}_{{\rm dd}}^{r}$. 

The Dyson equation for the dot Green function is written as
\begin{eqnarray}
\hat{G}_{{\rm dd}}^{r} = \hat{g}_{{\rm d}}^{r} + \hat{G}_{{\rm dd}}^{r} \hat{\Sigma}_{{\rm T}}^{r} \hat{g}_{{\rm d}}^{r} + \hat{D}^{r} \hat{U} \hat{g}_{{\rm d}}^{r}, \label{dyson eq}
\end{eqnarray}
where $\hat{U} =U \hat{\sigma}_{{\rm z}}$, and 
\begin{eqnarray}
\hat{g}_{{\rm d}}^{r}= \left[ \hat{I} \omega  - \hat{\sigma}_{{\rm z}} \epsilon_{{\rm d}} \right]^{-1},
\end{eqnarray}
\begin{eqnarray}
\hat{\Sigma}_{{\rm T}}^{r}=\hat{T}\left( \hat{G}_{{\rm LL}}^{(0) r} + \hat{G}_{{\rm RR}}^{(0) r} + \hat{G}_{{\rm RL}}^{(0) r} + \hat{G}_{{\rm LR}}^{(0) r} \right)\hat{T}.
\end{eqnarray}
Here $\hat{D}^{r}(\omega)=\langle \langle \psi_{{\rm d}} ;\varphi_{{\rm d}}^\dagger \rangle \rangle_{\omega}$ is the higher order Green function with
\begin{eqnarray}
\varphi_{{\rm d}}=\left( 
\begin{array}{ccc}
 d_{\uparrow} n_{\downarrow}  \\
 n_{\uparrow} d_{\downarrow}^{\dagger}  \\
\end{array} 
\right).
\label{Dyson}
\end{eqnarray}
In order to solve Eq.~(\ref{dyson eq}), we decouple $\hat{D}$ as
\begin{eqnarray}
\langle \langle \psi_{{\rm d}} ;\varphi_{{\rm d}}^\dagger \rangle \rangle_{\omega}
\approx
\langle \langle \psi_{{\rm d}} ;\psi_{{\rm d}}^\dagger \rangle \rangle_{\omega}
\left(
\begin{array}{ccc}
\langle n_{\downarrow} \rangle &
-\langle d_{\downarrow} d_{\uparrow} \rangle \\
\langle d_{\uparrow}^{\dagger} d_{\downarrow}^{\dagger} \rangle &
\langle n_{\uparrow} \rangle \\
\end{array}
\right), \label{HFD}
\end{eqnarray}
where the matrix elements are given by solving the usual self-consistent equations
\begin{eqnarray}
\langle n_{\uparrow} \rangle = \frac{1}{2 \pi i} \int {\rm d} \omega \left(\hat{G}_{{\rm dd}}^{<}(\omega) \right)_{11}&,
\\
\langle d_{\uparrow}^{\dagger} d_{\downarrow}^{\dagger} \rangle = \frac{1}{2 \pi i} \int {\rm d} \omega \left(\hat{G}_{{\rm dd}}^{<}(\omega) \right)_{21}&,
\end{eqnarray}
and the other elements obey similar equations.
The decoupling Eq.~(\ref{HFD}) corresponds to the usual HFA.
Substituting Eq.~(\ref{HFD}) into (\ref{dyson eq}), the dot Green function is obtained as
\begin{eqnarray} \label{dot green(0)}
\hat{G}_{{\rm dd}}^{r} = \left[ \left( \hat{g}_{\tilde{{\rm d}}}^{r} \right )^{-1} - \hat{\Sigma}_{{\rm T}}^{r} \right]^{-1},
\end{eqnarray}
where $\hat{g}_{\tilde{{\rm d}}}^{r}$ is the dressed Green function which is expressed as
\begin{eqnarray}
\left( \hat{g}_{\tilde{{\rm d}}}^{r} \right)^{-1} =
\left(
\begin{array}{ccc}
\omega - \epsilon_{{\rm d}} - U \langle n_{\downarrow} \rangle &
-U \langle d_{\downarrow} d_{\uparrow} \rangle \\
-U \langle d_{\uparrow}^{\dagger} d_{\downarrow}^{\dagger} \rangle &
\omega + \epsilon_{{\rm d}} + U \langle n_{\uparrow} \rangle \\
\end{array}
\right).
\end{eqnarray}
Then, we can finally estimate the Josephson current.
\begin{figure}[t]
\begin{center}
\includegraphics[width=64mm]{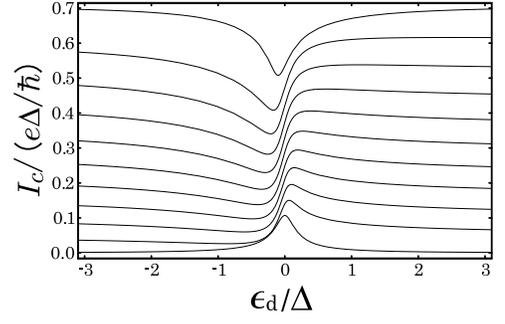}
\caption{\label{fig2}The critical currents in the non-interacting case ($U=0$) are plotted against the QD energy level $\epsilon_{{\rm d}}$. The transmission probability $\tau=4 \zeta/(1+\zeta)^{2}$ is changed from 0 (bottom) to 1 (top) with a step 0.1. The coupling strength is chosen to be $\Gamma/\Delta=0.1$, which is adapted to all figures below.}
\end{center}
\end{figure}
\begin{figure}[t]
\begin{center}
\includegraphics[width=64mm]{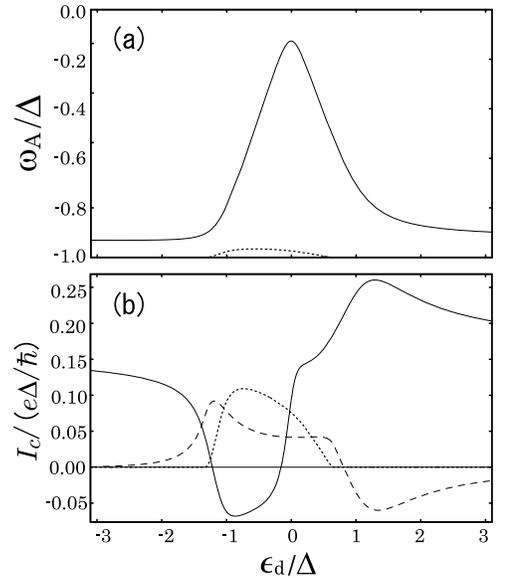}
\caption{\label{fig3}(a) ABSs below the Fermi level for $\tau=0.3$. The full (dotted) line corresponds to the primary (secondary) ABS, i.e., $\omega_{{\rm A}}=\omega_{1}~(\omega_{2})$. (b) Current contributions from each of the ABSs and the continuous spectrum for $\tau=0.3$. The dashed line indicates the continuous spectrum contribution.}
\end{center}
\end{figure}
\section{\label{3}RESULTS}
The Josephson current is obtained as a function of the phase difference $\theta=\theta_{{\rm L}}-\theta_{{\rm R}}$ and the QD energy level $\epsilon_{{\rm d}}$. We define the coupling strength between the QD and each of the superconductors by $\Gamma=\pi \nu t^{2}$. 
The direct tunneling across the insulating barrier is characterized by $\zeta=\pi^{2} \nu^{2} w^{2}$. Because the current-phase relation is roughly given by $I=I_{{\rm c}} \sin \theta$, for simplicity, we determine the Josephson critical current by $I_{{\rm c}} \equiv I(\theta=\pi/2)$ in this paper. All the calculations below are done assuming that the system is at the absolute zero of temperature. We set $\Gamma/\Delta=0.1$ throughout this paper.

The Josephson current consists of the two contributions; one from the discrete Andreev bound states (ABSs) inside the gap $\left| \omega \right| < \Delta$ and the other from the continuous spectrum outside the gap $\left| \omega \right| > \Delta$. The ABSs $\omega_{{\rm A}}$ are determined as the poles of the dot Green function, whereas the poles of the other Green functions such as $\hat{G}_{\beta \bar{\beta}}^{(0) r}$ do not contribute to the current.
\subsection{\label{A}Non-interacting case ($U=0$)}
We show the dependence of the critical current on the QD energy level in Fig.~\ref{fig2}. These line shapes remind us of the Fano line shapes which are seen in the differential conductance of normal metal systems. Far from the resonance ($|\epsilon_{{\rm d}}| \rightarrow \infty$), the currents asymptotically go to a certain value, which is given by
\begin{eqnarray}
I_{0}=\frac{e \Delta}{\hbar} \frac{\tau \sin \theta}{2 \sqrt{\mathstrut1 - \tau \sin^{2} \frac{\theta}{2}}},
\end{eqnarray}
where $\tau=4 \zeta/(1+\zeta)^{2}$ is the transmission probability through the direct channel.

The ABSs below the Fermi level are shown in Fig.~\ref{fig3}(a). One can see that there are the two ABSs, which we call ``primary ABS'' $\omega_1$ and ``secondary ABS" $\omega_2$ respectively. The secondary ABS is present in the region
\begin{eqnarray}
-\frac{\Gamma}{\sqrt[]{\mathstrut \zeta}}-\Delta<\epsilon_{{\rm d}}<-\frac{\Gamma}{\sqrt[]{\mathstrut \zeta}}+\Delta,
\end{eqnarray}
and disappears for $\zeta=0$, i.e., in the S/QD/S junction. We see that the ABSs represent the assymetric dependence on the QD energy level being affected by the interference. Figure~\ref{fig3}(b) shows the current contributions from each of the ABSs and the continuous spectrum outside the gap. 
It should be noticed that the contribution from the primary ABS becomes negative at a certain region, whereas the sum of the three is always positive. 

In the previous work, it was claimed that $0$-$\pi$ transition occurred around the dip even in the case $U=0$~\cite{senko}. We however cannot see the transition in the whole parameter regime. This difference is caused by the fact that the contribution from the continuous spectrum was not taken into account. It is known that the contribution from the continuous spectrum is not zero in S/QD/S junctions~\cite{SDS8,SDS9}. If we choose the appropriate parameter values ( $\Gamma/\Delta=4, \tau=0, 0.6, 1$ ), the contribution from the primary ABS shows almost the same behavior as Fig.~5(a) in Ref.~\onlinecite{senko}.

\subsection{\label{B}Interacting case ($U \neq 0$)}
\begin{figure}[t]
\begin{center}
\includegraphics[width=64mm]{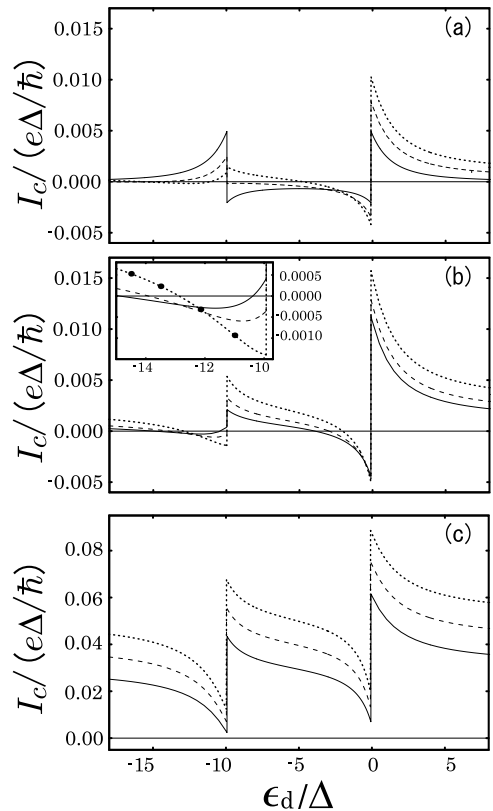}
\caption{\label{fig4} Critical currents as a function of the QD energy level at $U/\Delta=10$ for (a) $\tau=0$ (full line), $\tau=0.0005$ (dashed line), $\tau=0.0015$ (dotted line) (b) $\tau=0.002$ (full line), $\tau=0.003$ (dashed line), $\tau=0.005$ (dotted line) and (c) $\tau=0.06$ (full line), $\tau=0.08$ (dashed line), $\tau=0.1$ (dotted line). The inset of (b) is an extended figure around the resonance $\epsilon_{{\rm d}}=-U$.}
\end{center}
\end{figure}
\begin{figure}[t]
\begin{center}
\includegraphics[width=64mm]{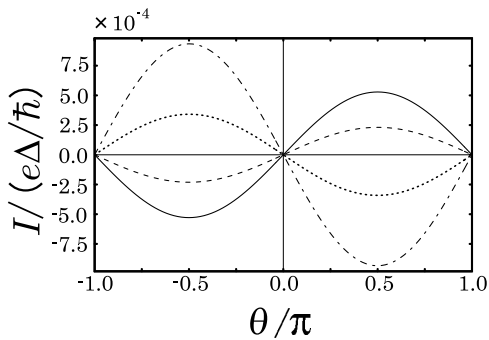}
\caption{\label{fig5} Current-phase relation at $U/\Delta=10$ for different $\epsilon_{{\rm d}}$ which are marked by the filled circles in the inset of Fig.~\ref{fig4} (b). Explicitly $\epsilon_{{\rm d}}/\Delta=-14.5$ (full line), $\epsilon_{{\rm d}}/\Delta=-13.5$ (dashed line), $\epsilon_{{\rm d}}/\Delta=-12.1$ (dotted line), $\epsilon_{{\rm d}}/\Delta=-11$ (dot-dashed line).}
\end{center}
\end{figure}
\begin{figure}[t]
\begin{center}
\includegraphics[width=64mm]{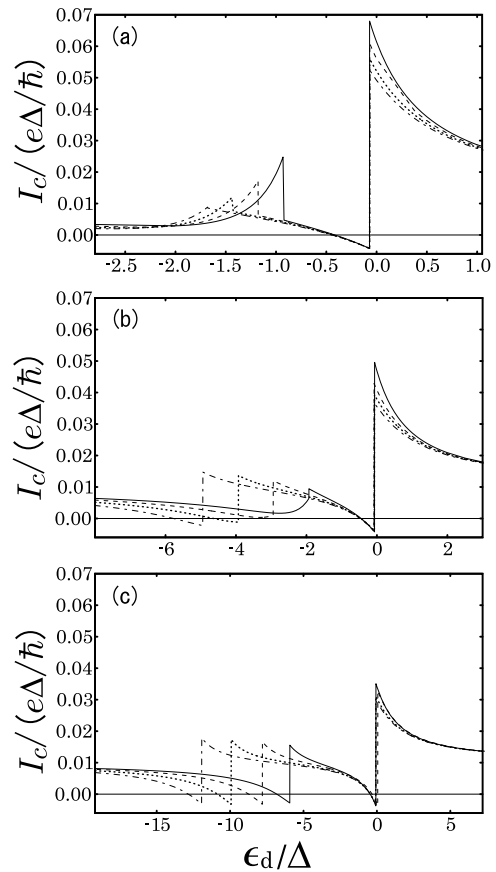}
\caption{\label{fig6} Critical currents as a function of the QD energy level for $\tau=0.02$ with (a) $U/\Delta=1$ (full line), $U/\Delta=1.25$ (dashed line), $U/\Delta=1.5$ (dotted line), $U/\Delta=1.75$ (dot-dashed line) (b) $U/\Delta=2$ (full line), $U/\Delta=3$ (dashed line), $U/\Delta=4$ (dotted line), $U/\Delta=5$ (dot-dashed line) and (c) $U/\Delta=6$ (full line), $U/\Delta=8$ (dashed line), $U/\Delta=10$ (dotted line), $U/\Delta=12$ (dot-dashed line)}
\end{center}
\end{figure}
We show in Fig.~\ref{fig4} the critical currents as a function of the QD energy level at $U/\Delta=10$ for different values of $\tau$. The full line in Fig.~\ref{fig4}(a), which is for $\tau=0$, represents the current flowing through the S/QD/S junction. As is verified by the previous theoretical~\cite{pi HF1,pi HF2} and experimental~\cite{pi2,SDS7} work, a negative critical current (i.e., $\pi$-junction) is found in the magnetic region $-U< \epsilon_{{\rm d}} < 0$, in which the average electron number in the QD is close to unity. 

With increasing $\tau$, the peak structure gradually changes because of the interference effect (Fano effect).
In Figs.~\ref{fig4}(a) and (b), focusing on the region $\epsilon_{{\rm d}}<-U$, the critical current near the resonance $\epsilon_{{\rm d}}=-U$ decreases as $\tau$ increases. Then, it becomes negative at a certain value of $\tau$, which means that $0$-$\pi$ transition occurs. We can clearly see in Fig.~\ref{fig5} that the current-phase relation changes from $I=|I_{{\rm c}}| \sin \theta$ to $I=-|I_{{\rm c}}| \sin \theta$ by sweeping $\epsilon_{{\rm d}}$ across the transition point.
Note that this $0$-$\pi$ transition occurs in the non-magnetic region $\epsilon_{{\rm d}}<-U$, where the average electron number is close to two. Thus, the transition is induced not by the same mechanism as that in S/QD/S junctions. 
We find in Fig.~\ref{fig4}(c) that the critical current is always positive for large $\tau$.

Figure~\ref{fig6} shows the critical currents against the QD energy level for different interaction strengths $U/\Delta$. For small $U/\Delta$, the peak structure around the resonance $\epsilon_{{\rm d}}=-U$ drastically changes with $U/\Delta$. Meanwhile, for large $U/\Delta$, the peak structure does not depend on the value of $U/\Delta$ (it just shifts the position of the resonance.).
Here the negative critical current in the dip region of the resonance $\epsilon_{{\rm d}}=-U$ is found again for large $U/\Delta$. 

As is shown above, the Fano effect induces $0$-$\pi$ transition with large $U/\Delta$ and appropriate values of $\tau$. Almost all the behaviors shown in this section are also seen in the results obtained by the Hubbard-I approximation~\cite{Hubbard}, which we show in the Appendix.
\section{\label{4}Perturbative approach}
It is important to find the physical origin of the $0$-$\pi$ transition and the parameter regime where the system is a $\pi$-junction.
To do this, we recalculate the Josephson current by a simple perturbation theory with respect to the electron tunneling.
The Josephson current flowing through our system has three components, $I=I_{{\rm i}}+I_{{\rm QD}}+I_{{\rm int}}$. Here $I_{{\rm i}}$ and $I_{{\rm QD}}$ are the non-interference components associated with the path through the insulating barrier and the QD respectively, and $I_{{\rm int}}$ is the interference component. They are expanded as $I_{{\rm i}}=I_{w^{2}}+I_{w^{4}}+ \cdot \cdot \cdot$, $I_{{\rm QD}}=I_{t^{4}}+I_{t^{6}}+ \cdot \cdot \cdot$, and $I_{{\rm int}}=I_{wt^{2}}+I_{w^{2}t^{2}}+ \cdot \cdot \cdot$, where $I_{w^{n}t^{m}}$ is the term which is proportional to $w^{n}t^{m}$.
We calculate the lowest order terms of each of the components.
Here we show the results only for $\epsilon_{{\rm d}}<-U$:
\begin{eqnarray}
\label{Iww}
&&I_{w^{2},{\rm c}} = \frac{e \Delta}{\hbar} \left( \frac{w}{\Delta} \right)^{2} \sum_{k,k'} \frac{2 \Delta^{3}}{E_{k}E_{k'}(E_{k}+E_{k'})}, \\ \nonumber \label{Itttt} 
&&I_{t^{4},{\rm c}} = \frac{e \Delta}{\hbar} \left( \frac{t}{\Delta} \right)^{4} \sum_{k,k'} \frac{2 \Delta^{5}}{E_{k} E_{k'} (E_{k} - \epsilon_{{\rm d}} - U) (E_{k'} - \epsilon_{{\rm d}} - U)} \nonumber \\
&&~~~~~~~~~~~~~~\times  \left( \frac{1}{E_{k}+E_{k'}} - \frac{2}{2 \epsilon_{{\rm d}} + U} \right), \\ \nonumber 
&&I_{wt^{2},{\rm c}} = -\frac{e \Delta}{\hbar} \left( \frac{w}{\Delta} \right) \left( \frac{t}{\Delta} \right)^{2} \sum_{k,k'} \frac{2 \Delta^{4}}{E_{k} E_{k'} (E_{k'} - \epsilon_{{\rm d}} - U)} \nonumber \\
&&~~~~~~~~~~~~~~\times  \left( \frac{2}{E_{k}+E_{k'}} + \frac{1}{E_{k} - \epsilon_{{\rm d}} - U} \right), \label{Iwtt}
\end{eqnarray}
where $I_{w^{n}t^{m},{\rm c}}=I_{w^{n}t^{m}}|_{\theta=\pi/2}$, and $E_{k}=\sqrt{\mathstrut \xi_{k}^{2}+\Delta^{2}}$.
One can see that only the term of the interference component is negative. It indicates that the interference effect can induce $0$-$\pi$ transition when $|I_{wt^{2},{\rm c}}|$ becomes large enough.
In fact, $I_{wt^{2},{\rm c}}$ changes its sign at the resonances and the particle-hole symmetric point $\epsilon_{{\rm d}}=-U/2$;
\begin{eqnarray}
I_{wt^{2},{\rm c}}\left\{ \begin{array}{ll}
<0 & (\epsilon_{{\rm d}}<-U) \\
>0 & (-U<\epsilon_{{\rm d}}<-U/2) \\
<0 & (-U/2<\epsilon_{{\rm d}}<0) \\
>0 & (0<\epsilon_{{\rm d}}) \\
\end{array} \right. .
\end{eqnarray}
This sign change leads to the characteristic peak-dip structure of the Fano resonance.

\begin{figure}[t]
\begin{center}
\includegraphics[width=75mm]{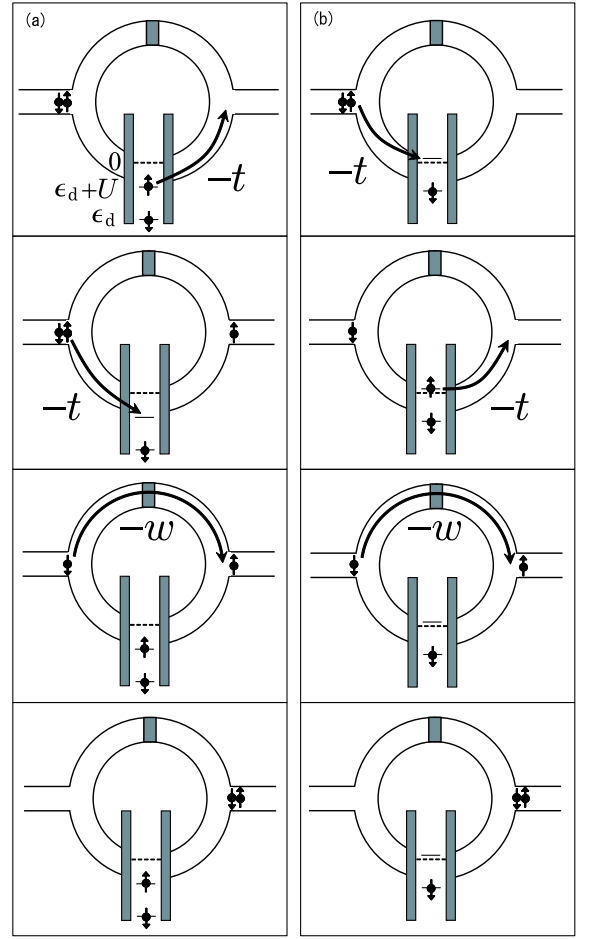}
\caption{\label{fig7} Schematic representation of tunneling processes which contribute to $I_{wt^{2}}$ for (a) $\epsilon_{{\rm d}}<-U$, and (b) $-U<\epsilon_{{\rm d}}<-U/2$. Dashed line in the energy diagrams of the QD means the Fermi level chosen as the zero of energy.}
\end{center}
\end{figure}

The term $I_{wt^{2},{\rm c}}$ originates from the tunneling processes in which the two electrons of a Cooper pair take the two different shortest paths. Two of such tunneling processes are schematically represented in Fig.~\ref{fig7}.
In each of the tunneling processes, the two electrons acquire the different additional phases $\phi_{{\rm QD}}$ and $\phi_{{\rm i}}$, respectively.
Then, the phase of the transported Cooper pair is shifted by $\phi=\phi_{{\rm i}}+\phi_{{\rm QD}}$.
The negative sign of $I_{wt^{2},{\rm c}}$ is attributed to this phase shift.

To examine what value $\phi$ takes in each of the tunneling processes, let us consider two of them.
Figure~\ref{fig7}(a) shows one for the case $\epsilon_{{\rm d}}<-U$, where the QD energy level is almost doubly occupied. In this case, to transfer an electron from the left to right superconductor through the QD, first an electron in the QD must tunnel out to the right superconductor. Then an electron in the left superconductor tunnels into the QD.
Meanwhile in the case $-U<\epsilon_{{\rm d}}<-U/2$ as in Fig.~\ref{fig7}(b), it is energetically favorable that first an electron tunnels into the QD.
Comparing these two tunneling processes, we see that the order of the elementary tunneling events is permuted.
Because of the anticommutation relation of electron operators, this permutation results in the $\pi$-difference of the additional phase $\phi_{{\rm QD}}$.

Calculating the tunneling amplitudes actually, one finds that $\phi_{{\rm QD}}=0$ ($\pi$) for the tunneling process represented in Fig.~\ref{fig7}(a) (Fig.~\ref{fig7}(b)).
On the other hand, the electrons tunneling through the insulating barrier get the additional phase $\phi_{{\rm i}}={\rm Arg}(-w)=\pi$.
As a result, the phase shift of the transported Cooper pair is given by $\phi=\pi$ ($0$) for the tunneling process shown in Fig.~\ref{fig7}(a) (Fig.~\ref{fig7}(b)).

The phase shift $\phi=\pi$ makes the tunneling process contribute to the critical current negatively.
For $\epsilon_{{\rm d}}<-U$, $\phi=\pi$ in all the tunneling processes which contribute to $I_{wt^{2}}$.
This is the main origin of the Fano-induced $0$-$\pi$ transition in our system.
Simply put, the essential origin of this transition is that the two electrons of a Cooper pair can take different paths and acquire additional phases the sum of which equals $\pi$.
If one changes the sign of $w$, the $\epsilon_{{\rm d}}$ dependence of the Josephson current is converted as $I(\epsilon_{{\rm d}}) \rightarrow I(-\epsilon_{{\rm d}}-U)$ because of the particle-hole symmetry.
However, one can always find the negative critical current in a non-magnetic region independently of the sign of $w$.

In the limit $\epsilon_{{\rm d}} \rightarrow -U-0^{+}$, where $|I_{wt^{2},{\rm c}}|$ has a maximum value, we can perform the summations in Eqs.~(\ref{Iww}) - (\ref{Iwtt}) analytically;
\begin{eqnarray}
&&I_{w^{2},{\rm c}} = \frac{e \Delta}{\hbar} 2 \zeta, \\  
&&I_{t^{4},{\rm c}} = \frac{e \Delta}{\hbar} \frac{2\Gamma^{2}}{\Delta^{2}} \left( \frac{4}{\pi}-1+\frac{2 \Delta}{U}\right), \label{Itttt2} \\
&&I_{wt^{2},{\rm c}} = -\frac{e \Delta}{\hbar} \frac{4~\sqrt[]{\mathstrut \zeta}~\Gamma}{\Delta}.
\end{eqnarray}
Substituting these expressions to $I_{c} \approx I_{w^{2},{\rm c}}+I_{t^{4},{\rm c}}+I_{wt^{2},{\rm c}} < 0$, we estimate the conditions for the negative critical current at $\epsilon_{{\rm d}} \rightarrow -U-0^{+}$;
\begin{eqnarray}
\label{condition1}
&&\zeta_{{\rm c}}^{-} < \zeta < \zeta_{{\rm c}}^{+}, \\
\label{condition2}
&&~~~U > U_{{\rm c}},
\end{eqnarray}
where 
\begin{eqnarray}
\label{xic}
\zeta_{{\rm c}}^{\pm} &=& \left[ 1 \pm \sqrt[]{\mathstrut 2 \left( 1 - \frac{2}{\pi} \right)-\frac{2 \Delta}{U}} \right]^{2} \left( \frac{\Gamma}{\Delta} \right)^{2},
 \\
U_{{\rm c}} &=& \frac{\Delta}{1 - 2/\pi} \approx 2.75 \Delta.
\end{eqnarray}
From Eqs.(\ref{condition1}) and (\ref{condition2}), we see that the negative critical current is possible only for large $U$ and the appropriate values of $\zeta$.
It is consistent with the results obtained by the HFA shown in the previous section. 
The critical values of $\tau$ for $U/\Delta=10$ and $\Gamma/\Delta=0.1$ are given by $\tau_{{\rm c}}^{\pm} \equiv 4\zeta_{{\rm c}}^{\pm}/(1+\zeta_{{\rm c}}^{\pm})^{2} \approx 0.003,~0.112$. These are somewhat larger than the values $\tau_{{\rm HFA},{\rm c}}^{\pm} \approx 0.002, 0.05$ estimated from the results of the HFA. 
Eq.~(\ref{xic}) tells us that $w/\Delta \sim (t/\Delta)^{2}$ is necessary for the negative critical current. In this regime, the interference and non-interference components are the same order, and the equation $I = I_{w^{2}}+I_{t^{4}}+I_{wt^{2}}$ is correct within the second order of $\Gamma/\Delta$.

\section{\label{5}SUMMARY}
We have investigated the Fano effect in a Josephson junction including a QD in the weak coupling regime $T_{{\rm K}} \ll \Delta$. To treat the Coulomb interaction inside the QD, we employ the HFA. The interference effect between the direct and resonant tunneling processes causes the characteristic Fano line shape. Furthermore, it is found that our system behaves as a $\pi$-junction even in the non-magnetic region $\epsilon_{{\rm d}}<-U$ for large $U/\Delta$.
These results suggest that $0$-$\pi$ transition is induced by the cooperation of the Fano effect and the Coulomb interaction inside the QD.

A superconducting quantum interference device (SQUID) including carbon nanotube quantum dots was already achieved in the recent experiment~\cite{SDS7}. We expect that our theory will be experimentally verified using the similar system in the near future.

\acknowledgments
We are grateful to K. Kamide and D. Yamamoto for valuable comments and discussions. 
\begin{appendix}
\section{Hubbard-I approximation}
In this Appendix, we calculate the Josephson current by the Hubbard-I approximation~\cite{Hubbard} (HIA) generalized to the superconducting case.
By using the HIA, one can take into account the correlation $\langle \delta n_{\uparrow} \delta n_{\downarrow} \rangle$ beyond the simple HFA.
Since this correlation largely contributes to the charge fluctuation $\sqrt[]{\mathstrut \langle (\delta n_{\uparrow} + \delta n_{\downarrow})^{2} \rangle}$, we believe that the HIA is more reliable at least near the resonances where the charge fluctuation plays an important role.

First we derive the equation of motion for $D$ in Eq.~(\ref{dyson eq}), and then decouple the higher order Green functions which arise from $\langle \langle \psi_{{\rm d}}; [{\cal H}, \varphi_{{\rm d}}^{\dagger}] \rangle \rangle_{\omega}$. After that, we get the following equation;
\begin{eqnarray}
\langle \langle \psi_{{\rm d}} ;\varphi_{{\rm d}}^\dagger \rangle \rangle_{\omega}
\approx
\hat{\Lambda}_{+} \hat{g}_{{\rm u}}^{r} +
\sum_{\beta} \langle \langle \psi_{{\rm d}} ;\psi_{\beta}^{\dagger} \rangle \rangle_{\omega} \hat{\Lambda}_{-} \hat{T} \hat{g}_{{\rm u}}^{r}
\nonumber\\
+ \langle \langle \psi_{{\rm d}} ;\psi_{{\rm d}}^{\dagger} \rangle \rangle_{\omega} \hat{\Lambda} \hat{T} \hat{g}_{{\rm u}}^{r}
, 
\label{EOM for D}
\end{eqnarray}
where
\begin{eqnarray}
&&\hat{g}_{{\rm u}}^{r} = \left[ \hat{I} \omega - \hat{\sigma}_{{\rm z}} (\epsilon_{{\rm d}} + U) \right]^{-1},
\\
&&\hat{\Lambda}_{\pm} = 
\left(
\begin{array}{ccc}
\langle n_{\downarrow} \rangle &
\mp \langle d_{\downarrow} d_{\uparrow} \rangle \\
\pm \langle d_{\uparrow}^{\dagger} d_{\downarrow}^{\dagger} \rangle &
\langle n_{\uparrow} \rangle \\
\end{array}
\right),
\\
&&\hat{\Lambda} = 
\sum_{\beta}
\left(
\begin{array}{ccc}
\langle c_{\beta,\downarrow}^\dagger d_{\downarrow} \rangle -
\langle d_{\downarrow}^\dagger c_{\beta,\downarrow} \rangle  &
\langle d_{\uparrow} c_{\beta,\downarrow} \rangle  +
\langle c_{\beta,\uparrow} d_{\downarrow} \rangle  \\
\langle d_{\uparrow}^\dagger c_{\beta,\downarrow}^{\dagger} \rangle +
\langle c_{\beta,\uparrow}^\dagger d_{\downarrow}^{\dagger} \rangle  &
\langle d_{\uparrow}^\dagger c_{\beta,\uparrow} \rangle -
\langle c_{\beta,\uparrow}^\dagger d_{\uparrow} \rangle \\
\end{array}
\right). \nonumber \\ \label{eq30}
\end{eqnarray}
Equation~(\ref{eq30}) seems complex to treat for a first look. However, assuming that the current through the QD is uniform, one can see that the following relation holds
\begin{eqnarray}
\langle \frac{d}{dt} n_{\sigma} \rangle
=
-i t \sum_{\beta} \left(\langle c_{\beta,\sigma}^\dagger d_{\sigma} \rangle -
\langle d_{\sigma}^\dagger c_{\beta,\sigma} \rangle \right) = 0. 
\label{AS1}
\end{eqnarray}
In addition, we assume that one can disregard the time derivative of the pair amplitude inside the QD, i.e.,
\begin{eqnarray}
\langle 
\frac{d}{dt} ( d_{\uparrow}^{\dagger} d_{\downarrow}^{\dagger} ) \rangle 
=
(2\epsilon_{{\rm d}}+U)\langle d_{\uparrow}^{\dagger} d_{\downarrow}^{\dagger}
\rangle
~~~~~~~~~~~~~~~~~~~~~&&
\nonumber \\
-t\sum_{\beta}
\Big(
\langle c_{\beta,\uparrow}^{\dagger} d_{\downarrow}^{\dagger} \rangle
+ \langle d_{\uparrow}^{\dagger} c_{\beta,\downarrow}^{\dagger} \rangle 
\Big)=0, &&
\label{AS2}
\end{eqnarray}
which guarantees the condition $(\hat{G}_{{\rm dd}}^{r})^{\dagger}=\hat{G}_{{\rm dd}}^{a}$. Thus Eq.~(\ref{eq30}) is simplified as
\begin{eqnarray}
\hat{\Lambda} = 
\frac{(2 \epsilon_{{\rm d}} + U)}{t}
\left(
\begin{array}{ccc}
0 &
-\langle d_{\downarrow} d_{\uparrow} \rangle \\
\langle d_{\uparrow}^\dagger d_{\downarrow}^\dagger \rangle &
0 \\
\end{array}
\right).
\label{simplelambsa}
\end{eqnarray}
The dot Green function is formally obtained by substituting Eq.~(\ref{EOM for D}) to (\ref{dyson eq}) as 
\begin{eqnarray}
\hat{G}_{{\rm dd}}^{r}
=
\left[
\left( \hat{g}_{{\rm u}}^{r} \right )^{-1} + \hat{\Lambda}_{+} \hat{U}
\right]~~~~~~~~~~~~~~~~~~~~~~~~~~~~~~~~~~~ &&
\nonumber \\
\times \left[
\left( \hat{g}_{{\rm u}}^{r}\hat{g}_{{\rm d}}^{r} \right )^{-1} 
- \hat{\Sigma}_{{\rm T}}^{r} \left( \left( \hat{g}_{{\rm u}}^{r} \right )^{-1} + \hat{U} \hat{\Lambda}_{-} \right) 
-
\hat{\Lambda} \hat{T} \hat{U}
\right]^{-1}.&& \nonumber \\
\end{eqnarray}
\begin{figure}[t]
\begin{center}
\includegraphics[width=65mm]{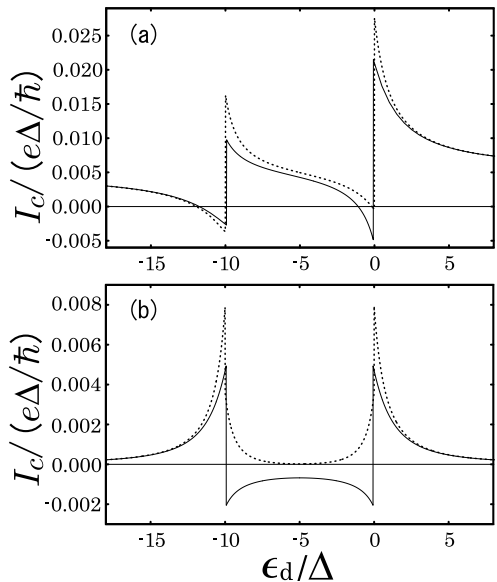}
\caption{\label{fig8} Plots of the critical currents versus the QD energy level  at $U/\Delta=10$ for (a) $\tau=0.01$ and (b) $\tau=0$. The full (dotted) lines represent the results obtained by the HFA (HIA).}
\end{center}
\end{figure}
From this dot Green function, we estimate the Josephson current in the same way as in Sec.~I\hspace{-.1em}I.

In Fig.~\ref{fig8}, we show the results of the HIA together with the ones of the HFA. In the case $\tau \neq 0$, the two results show almost the same behavior, and the negative critical current in the region $\epsilon_{{\rm d}}<-U$ is found in both the results.
Since, compared to the HIA, the charge fluctuation is underestimated in the HFA, the difference between the two results becomes large as $\epsilon_{{\rm d}}$ approaches the resonances.

Meanwhile, for $\tau=0$, the obvious difference is seen in the region $-U<\epsilon_{{\rm d}}<0$. That is, a negative critical current is not seen in the results of the HIA.
It means that the HIA fails to demonstrate the $\pi$-junction behavior in the S/QD/S junction. This is because, in the HIA, one overestimates the spin fluctuation which is considered to be small in this region. Hence, the spin doublet state is not well described. 
This discrepancy should be corrected, but this is out of the scope of this paper.
\end{appendix}

\newpage 

%

\end{document}